# Scaling of the Hall effects beyond the quantum resistance threshold in oxidized CoFeB.

A. Gerber and G. Kopnov

Raymond and Beverly Sackler Faculty of Exact Sciences

School of Physics and Astronomy

Tel Aviv University

Ramat Aviv, 69978 Tel Aviv, Israel

The ordinary and the extraordinary Hall effects were studied in gradually oxidized amorphous CoFeB ferromagnets over six orders of resistivity from the metallic to the strongly insulating regime. Polarity of the extraordinary Hall effect reverses, and the amplitude of both the ordinary and the extraordinary Hall effects increases quadratically with resistivity when resistance exceeds the quantum resistance threshold $h/2e^2$. The absolute value of the extraordinary Hall effect scales linearly with the ordinary one in the entire range over eight orders of magnitude between the metallic and the insulating states. The behavior differs qualitatively and quantitatively from theoretically predicted and experimentally known in other materials.

# Introduction.

Hall effect is the major source of information on a type and density of charge carriers. Regretfully, our knowledge of the phenomenon in strongly disordered conductors is quite poor as experimental data are scarce and sometimes controversial. The reason is essentially technical: for any practical arrangement of electric contacts, there is an unavoidable geometrical mismatch in positioning of transverse to current (Hall) probes. Since longitudinal electric field is usually much larger than the transverse one, a relatively small Hall signal is hidden on a background of the longitudinal resistance noise. Thus, practically no data are available on the insulating side of the metal-insulator transition for resistivity exceeding $0.1 - 1$ Ωcm. Absence of data does not reflect an absence of interest, as a number of puzzling phenomena, like an existence of the Hall insulator state [1-3], double reversal of Hall effect polarity [4, 5] and the giant Hall effect [6] were pointed out but remained unresolved. Availability of the extraordinary Hall effect (EHE) data in ferromagnetic systems is somewhat better, since the effect is usually much larger than the ordinary Hall effect and is easier to measure. An impression has been created that recent theoretical models [7 – 10] provide satisfactory interpretation to the entire existing data, excluding probably the giant EHE in granular ferromagnets [11]. Here we report on the experimental study of the ordinary and the extraordinary Hall effects in gradually oxidized amorphous CoFeB ferromagnets, where the effects appear to be remarkably large and measurable over six orders of resistivity from the metallic into the deep insulating regime. We found a number of unexpected and intriguing properties that include a crossover between the apparently Hall insulator state and the state with quadratically diverging Hall coefficient; a linear correlation between the ordinary Hall coefficient and the extraordinary Hall resistivity expanding from the metallic to the insulating state; and collapse of the accepted EHE scaling in the high resistivity limit. The crossover resistance where the unusual behavior starts is very close to the quantum resistance value.

**Experimental.**

Recently we reported [12] on development of partially oxidized amorphous CoFeB ferromagnetic films with resistivity variable between 100 μΩcm to beyond $10^2$ Ωcm, depending on degree of oxidation. Thin films with thickness between 5 and 50 nm were fabricated by reactive RF magnetron sputtering from $Co_{40}Fe_{40}B_{20}$ target (ACI Alloys Inc.) on rectangular 5x5mm$^2$ pieces of intrinsic GaAs substrate. Base pressure prior to deposition was about $2\times10^{-7}$ torr, whereas deposition took place at $5\times10^{-3}$ torr Ar atmosphere mixed with a controlled flow of either air or pure oxygen. Typical deposition rate was 0.1-0.2 nm/sec. Resistivity is very sensitive to the presence of air and increases sharply when air partial pressure goes above $10^{-3}$ torr, that is about 1:5 ratio with argon. Structural analyzes were done using X- ray diffraction, high-resolution transmission electron microscopy and TOF-SIMS. All samples deposited with and without air were found to be amorphous [12]. Resistance, magnetoresistance and Hall effect were measured using Van der Pauw protocol. The Hall effect and magnetization data discussed here were obtained at room temperature. The GaAs substrate we used was not conducting when tested by itself, and the sufficiently oxidized films were not conducting as well. We therefore excluded the possibility of current leakage and parallel conductance along the films and the substrate [13]. Magnetic characterization of the samples was done using SQUID magnetometer.

Due to the amorphous structure, we did not succeed to extract information on topology of the oxidized samples: whether the material is homogeneous or heterogeneous, forming amorphous metallic clusters embedded within amorphous insulating oxide. As will be discussed in the following, microscopic interpretation of the data might depend on this missing information.

**Results and discussion.**

Fig.1 presents the field dependence of Hall resistance $R_{xy}$ of two typical samples with lower (a) and higher (b) degrees of oxidation. Both samples are 10 nm thick; their resistivity are $11.5\times10^{-3}$ Ωcm (a) and 1.4 Ωcm (b). Hall resistance $R_{xy}$ in magnetic films can be presented as a superposition of the ordinary and the extraordinary Hall effects: $R_{xy} = V_{xy}/I = \rho_{xy}/t = (\rho_{OHE} B_z + \mu_0 R_s M_z)/t$, where $\rho_{xy}$ is Hall resistivity, $I$ – current, $t$ - thickness, $\rho_{OHE}$ and $R_s$ are the ordinary and the extraordinary Hall effect coefficients and $B_z$ and $M_z$ are the normal-to-plane projections of magnetic field induction and magnetization, respectively. EHE resistivity $\rho_{EHE}$ is defined as: $\rho_{EHE} = \mu_0 R_s M_{z,sat}$, where $M_{z,sat}$ is the saturated out-of-plane magnetization. $\rho_{EHE}$ is determined by extrapolation of the magnetically saturated high field linear portion of $R_{xy}(B)$ to zero field. The ordinary Hall effect coefficient $\rho_{OHE}$ is determined by the high field slope in magnetically saturated state as: $\rho_{OHE} = dR_{xy}/dB \cdot t$. $\rho_{OHE}$ is constant in field over at least 10 T range and is negative in all samples. The extraordinary Hall resistivity, dominant below 1 T, is negative in sample (a) and positive in the high resistivity samples (b).

Fig. 2 shows the ordinary Hall effect coefficient $\rho_{OHE}$ as a function of longitudinal resistivity $\rho$. Within the logarithmic accuracy $\rho_{OHE}$ increases slightly with increase of resistivity below $10^{-1}$ Ωcm, and grows sharply as: $\rho_{OHE} \propto \rho^2$ for $\rho > 10^{-1}$ Ωcm. $\rho_{OHE}$ exceeds $10^{-1}$ Ωcm at $\rho = 10^2$ Ωcm, which is six orders of magnitude higher than in bulk amorphous CoFeB. Samples sputtered in a pure Ar atmosphere are metallic with a positive resistivity temperature coefficient at room temperature. Oxidized samples are insulator-like with resistivity increasing with decreasing temperature. Transition between the metallic-like and the insulator-like temperature dependencies occurs at resistivity about $10^{-3}$ Ωcm, marked by the M-I arrow in the figure. Transition between the range with approximately constant Hall coefficient and the range in which $\rho_{OHE} \propto \rho^2$ takes place at about $10^{-1}$ Ωcm, well beyond the metal-insulator transition. The inset in Fig. 2 presents the resistivity temperature dependence of two samples: the first belonging to the constant Hall range and the second to the diverging one. Resistivity of both samples can be presented as: $\rho = \rho_0 exp\left(\frac{T_0}{T}\right)^{1/4}$ with $T_0$ increasing from 150 K for sample 1 with room temperature resistivity $10^{-2}$ Ωcm to about $10^7$ K for sample 2 with resistivity 3.2 Ωcm. Such temperature dependence of resistivity is consistent with the variable range hopping conductance model [14], but also with the temperature assisted conductance in granular media [15].

To emphasize an unusual character of the high resistivity Hall coefficient we plot in Fig. 3 the effective Hall mobility $\mu_{eff}$ as a function of sheet resistance. $\mu_{eff}$ was calculated as: $\mu_{eff} = \sigma/qn^* = \rho_{OHE}/\rho$, where $n^*$ is defined as an "effective" charge density. We use the terms "effective density" and "effective mobility", since the classical definition $\rho_{OHE} = 1/nq$, where $q$ is electric charge and n is a number of free electrons per unit volume, is not generally correct for hopping charge transport [16, 17]. The effective mobility has a unique V shape as a function of resistance: decreasing linearly with resistance in the "low" resistance range and increasing linearly at high resistances. As a rule, mobility decreases with increasing disorder and respectively increasing resistivity. Thus, a linear increase of mobility in the high resistance limit is remarkable. Notably, resistance of the crossover between the decreasing and increasing mobility ranges is close to the quantum resistance value $R_Q = h/2e^2 \approx 12.9$ kΩ.

Hall effect in the hopping conductance regime was calculated in a number of works [17-23]. It is commonly considered that in macroscopically homogeneous material the hopping Hall effect is related to self-interference of the electron wave function, propagating along different paths with at least three localization centers taken into account [17]. Mainly, it was the temperature variation of the effect that has been addressed. The power law relation between the Hall coefficient and longitudinal resistance was predicted to be $\rho_{OHE} \propto \rho^\gamma$, with the exponent $\gamma < 1$. Specifically, the value $\gamma = 0.35$ was predicted for conduction with a variable hopping length, and $\gamma = 0.5$ for conduction with a constant activation energy [21, 22]. On the other hand, an intriguing Hall insulator state was predicted [1] for 2D electron gas, in which resistance increases to infinity upon lowering the temperature, while the Hall coefficient remains finite and close to the classical value $\rho_{OHE} = 1/n_s e$, where $n_s$ is an areal density of electrons. Our data for $\rho < 10^{-1}$ Ωcm seem to meet the definition of the Hall insulator state with diverging resistance and approximately constant Hall coefficient, although resistivity increases due to oxidation and not by decreasing temperature. A different interpretation of the Hall coefficient independent on resistivity was proposed by Kharitonov and Efetov [24, 25] for heterogeneous granular materials with intergranular resistance smaller than the quantum resistance $R_Q$. The model assumes that Hall voltage is generated within grains only, and in absence of quantum effects, is given by the classical formula $\rho_{OHE} = 1/n^* e$, where $n^*$ differs from the carrier density $n$ inside the grains by a numerical coefficient determined by the shape of the grains and type of granular lattice. Resistivity, on the other hand, is determined by intergranular tunneling. Therefore, scaling between resistivity and Hall resistivity is absent. The predicted absence of scaling in this regime has been observed in granular $NiSiO_2$ mixtures [26]. To the best of our knowledge, the existing models of the hopping Hall effect in macroscopically homogeneous media predict no crossover phenomenon when resistance exceeds any critical value. Also, there is no Hall effect model for the heterogeneous granular material with intergranular resistance larger than $R_Q$.

We shift now to the extraordinary Hall effect. Recent models of EHE in multiband ferromagnetic metals with diluted impurities [7-9] predict three distinct scaling regimes as a function of conductivity. In the clean regime ($\sigma > 10^6$ $(\Omega cm)^{-1}$), the skew scattering mechanism ($\rho_{EHE} \propto \rho$, or $\sigma_{EHE} \propto \sigma$) is predicted to dominate. The extrinsic side jump and an intrinsic Berry phase mechanisms, both characterized by the ratio: $\rho_{EHE} \propto \rho^2$, (or $\sigma_{EHE} = const$), are expected to be dominant in the intermediate disorder regime ($\sigma \sim 10^4 - 10^6$ $(\Omega cm)^{-1}$). In the high disorder range ($\sigma < 10^4$ $(\Omega cm)^{-1}$), the intrinsic contribution is strongly decayed, resulting in a scaling relation $\sigma_{EHE} \propto \sigma^\gamma$ with $\gamma \sim 1.6$. The theory is based on the use of Bloch wave functions assuming a metallic conduction; hence its results are valid only for ferromagnetic metals in principle. EHE in insulating materials with phonon-assisted hopping conductance was treated in Ref. [10]. Here, EHE was calculated by considering hopping through triads of sites [17] along percolating clusters. Scaling $\sigma_{EHE} \propto \sigma^\gamma$ with $1.33 \leq \gamma \leq 1.76$ has been predicted for arbitrary thermally activated hopping processes including variable range hopping, short-range activation hopping or tunneling influenced by interactions in the Efros-Shklovskii regime. Thus, universal scaling in the form $\sigma_{EHE} \propto \sigma^\gamma$ with $\gamma \sim 1.6$ is anticipated for low conductivity materials regardless of whether their conductivity is metallic or thermally activated. Experimental data accumulated so far for different ferromagnets, including perovskite oxides, spinels and magnetic semiconductors [8] seem to be in a reasonable agreement with these theoretical predictions. Fig. 4 presents the absolute value of the extraordinary Hall conductivity in oxidized CoFeB films as a function of longitudinal conductivity. One can clearly distinguish two ranges: 1) $\sigma > 10$ $(\Omega cm)^{-1}$, where $\sigma_{EHE} \propto \sigma^2$, and 2) $\sigma < 10$ $(\Omega cm)^{-1}$, where $\sigma_{EHE} \approx const$. The seeming permanence of $\rho_{EHE}$ ($\sigma_{EHE} \propto \sigma^2$) is an artifact of the EHE conductivity presentation in logarithmic scale. $\rho_{EHE}$ magnitude increases slightly (factor of 1.8) in mildly oxidized samples when resistivity increases by two orders of magnitude between $6\times10^{-4}$ $\Omega cm$ and $6\times10^{-2}$ $\Omega cm$. With further increase of resistivity $\rho_{EHE}$ starts dropping, reverses polarity to positive and grows beyond 1 $\Omega cm$ when resistivity reaches $10^2$ $\Omega cm$. The onset of the polarity reversal was found in the sample with sheet resistance $1.2\times10^4$ $\Omega$, which is remarkably close to the quantum resistance value $R_Q = h/2e^2 \approx 12.9$ k$\Omega$. Thus, in the "high" conductivity range the EHE conductance follows $\sigma_{EHE} \propto \sigma^2$, which is not too far from the expected $\sigma_{EHE} \propto \sigma^{1.6}$. However, the "low" conductivity range, where EHE conductance reverses its polarity and is independent on conductivity, is in a striking conflict with any known to us model of EHE. The transition between the two ranges occurs when resistance exceeds the quantum resistance threshold.

Magnetization of a series of 50 nm thick samples as a function of their resistivity is shown in the inset of Fig. 4. Magnetization decreases with oxidation from about 600 emu/cm$^3$ in the non-oxidized sample and drops sharply when resistivity exceeds the same critical threshold. This indicates a clear correlation between the magnetization and the effective density of charge carriers.

Polarity of the Hall effects in the hopping conductance regime was considered in [Ref. 27] for hopping of holes between localized states in the impurity band of GaAsMn. In this case, $\sigma_{EHE}$ was shown to be proportional to the derivative of the density of states at the Fermi energy and, therefore expected to change sign as the Fermi level crosses the density-of-states maximum in the impurity band. The ordinary Hall coefficient was predicted to have the same sign everywhere in the impurity band. We find no evidence for correlation between the density of states maximum and crossing the quantum resistance value, therefore relevance of this model to our case is in question.

It is illuminating to compare the ordinary and the extraordinary Hall effects. Fig. 5 presents the absolute magnitude of EHE resistivity as a function of the ordinary Hall coefficient. The figure includes few polycrystalline non-oxidized and slightly oxidized samples fabricated using post-deposition annealing that exhibit Hall effects significantly smaller than amorphous samples. $\rho_{EHE}$ is a linear function of $\rho_{OHE}$ over almost eight orders of magnitude starting from the metallic samples and up to the strongly insulating ones. $R_Q$ marks the crossover threshold, where $\rho_{EHE}$ reverses its polarity and both $\rho_{EHE}$ and $\rho_{OHE}$ start increasing as $\rho^2$. The linear correlation between $\rho_{EHE}$ and $\rho_{OHE}$ is preserved in the entire range both below and above the threshold. Similar linear correlation between $\rho_{EHE}$ and $\rho_{OHE}$ has been reported in granular $NiSiO_2$ in the metallic and strongly coupled granular ranges [26]. Interpretation of such correlation within the classical single band conduction model would mean that the EHE conductivity is proportional to the charge carrier density, as predicted by Noziéres and Lewiner [28]. However, no such correlation was predicted and not even discussed for the hopping or temperature activated tunneling conductance.

The magnitude of the effects is remarkable by itself. Classical models predict the Hall coefficient in metals to depend on density of carriers and not on the mean free path, therefore predict no significant changes with resistivity. In granular percolating systems, as the metal concentration approaches the percolation threshold, the Hall coefficient is expected to scale together with conductivity and increase by about tenfold compared with a pure bulk metal. It was surprising when three to four orders enhancement of the extraordinary and the ordinary Hall coefficients have been observed in ferromagnetic granular mixtures $NiSiO_2$ [29], $FeSiO_2$ [30] and non-magnetic granular $Cu-SiO_2$ [6] and $Mo-SnO_2$ [31]. A local quantum interference theory was suggested [6, 32], in which presence of small insulating substructures along an infinite metallic cluster leads to profound wave scattering and interference, and thus causing significant reduction of the effective carrier density. The model is only valid at low temperatures when quantum corrections (weak localization / electron-electron interaction) are valid, and cannot explain the "giant Hall effect" at room temperature. Here, we find a huge enhancement by almost eight orders of magnitude in both Hall effects, surpassing by large the giant Hall effect in granular systems.

**Summary.**

We can summarize our main findings as following: 1) Amorphous ferromagnetic CoFeB films can be controllably modified by gradual oxidation between the metallic and the insulating states over six orders of resistivity. Both the ordinary and the extraordinary Hall effects are large enough to be measurable in the strongly insulating state. 2) There exists a critical resistance threshold beyond which three effects have been identified: polarity of the EHE reverses, and both the ordinary Hall coefficient and the extraordinary Hall resistivity amplitudes diverge with resistivity as $\rho^2$. 3) The critical threshold can be identified as the quantum resistance $h/2e^2$. 4) The EHE resistivity scales linearly with the ordinary Hall coefficient over eight orders of magnitude from the metallic into the strongly insulating state. 5) Both the EHE resistivity and the ordinary Hall coefficient in the oxidized state are huge, exceeding their values in the crystalline metallic state by almost eight orders of magnitude. The scope of the behavior is dramatically different both qualitatively and quantitatively from the experimentally known in other studied materials and from the theoretically predicted. Unfortunately, we were unable to establish the microscopic structure of this amorphous material. Whether it is homogeneous or heterogeneous, i.e. composed of amorphous ferromagnetic clusters embedded within amorphous insulating matrix is currently not known.  One wonders whether the phenomena are a general property of amorphous partially oxidized normal and ferromagnetic metals, and whether this type of materials with a huge magnetic field response and non-trivial effective mobility can be used for practical applications.

**Acknowledgements**

We acknowledge the financial support by the XIN center of Tel Aviv and Tsinghua Universities.

# References.

**Figure captions.**

Fig.1. Hall resistance $R_{xy}$ of two 10 nm thick samples with lower (a) and higher (b) degrees of oxidation as a function of applied magnetic fields. Resistivity is 11.5×10$^{-3}$ Ωcm (a) and 1.4 Ωcm (b).

Fig.2. The ordinary Hall effect coefficient as a function of longitudinal resistivity ρ. M-I indicates the metal-insulator transition. Inset: resistivity of two typical samples as a function of temperature, taken from Ref. 12. Straight lines are guide to the eye.

Fig.3. The effective mobility of a series of 50 nm thick samples as a function of sheet resistance. Straight lines are guide to the eye.

Fig.4. The absolute value of the EHE conductivity as a function of longitudinal conductivity. Straight lines are guide to the eye. Inset: magnetization of 50 nm thick samples as a function of resistivity.

Fig.5. The absolute value of the EHE resistivity as a function of the ordinary Hall coefficient for multiple series of different thickness. Open symbols correspond to the annealed polycrystalline samples. The slope of the straight line is 1. $R_Q$ marks the crossover threshold.

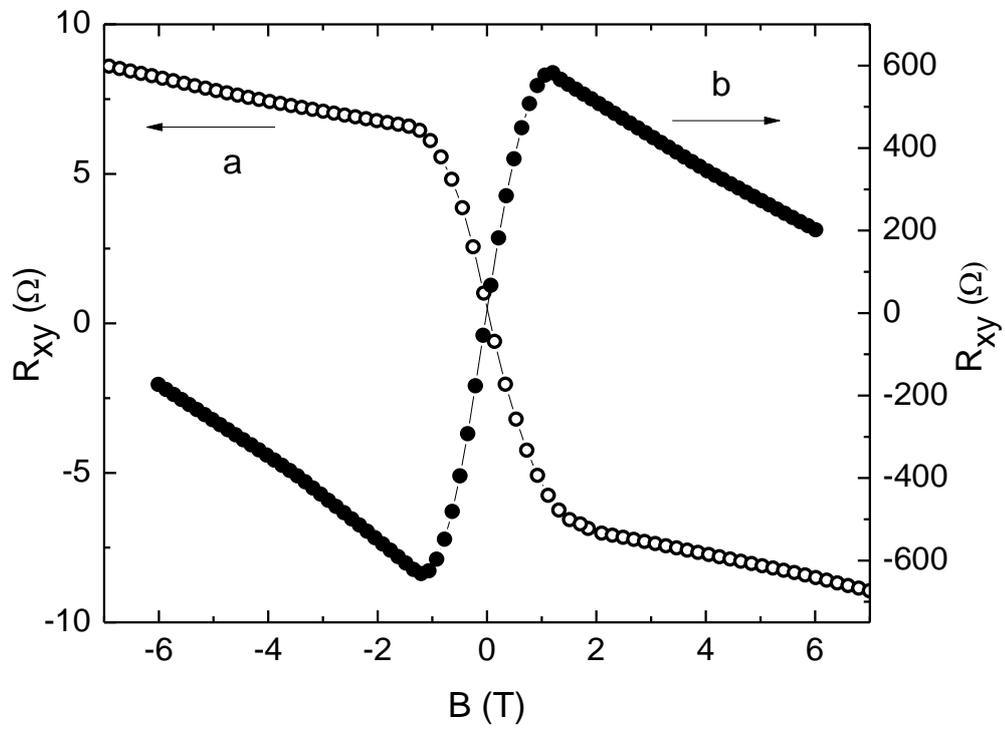

Fig.1

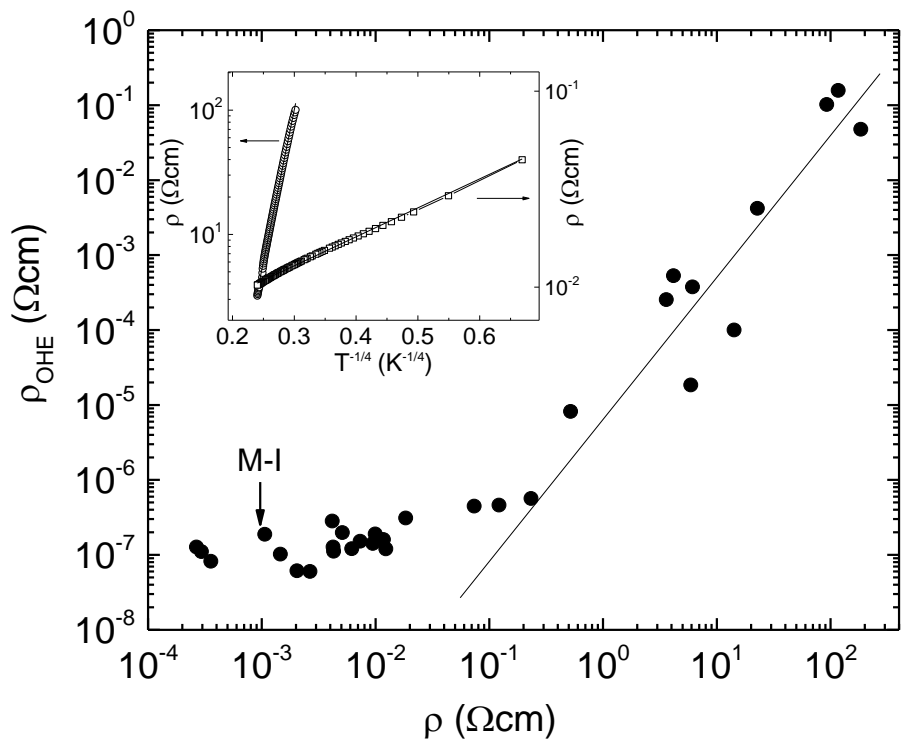

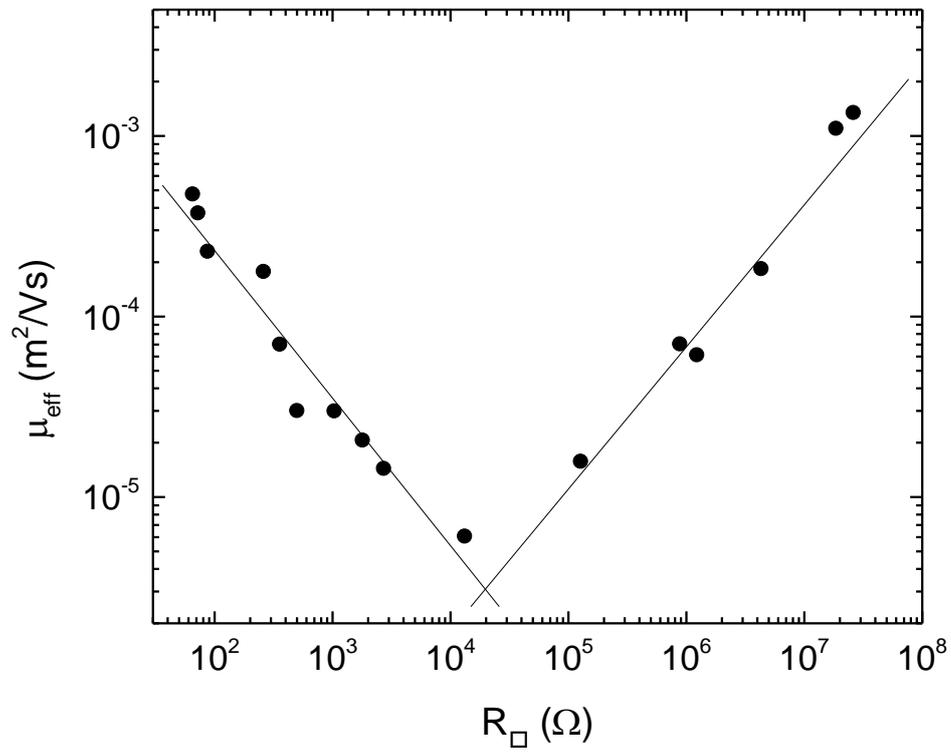

Fig. 3

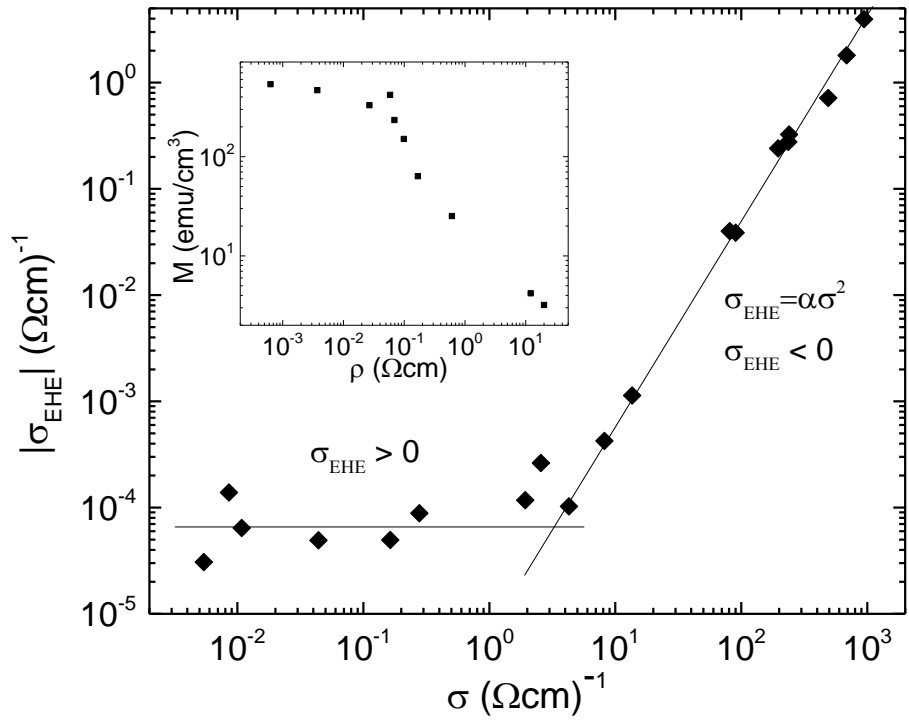

Fig. 4

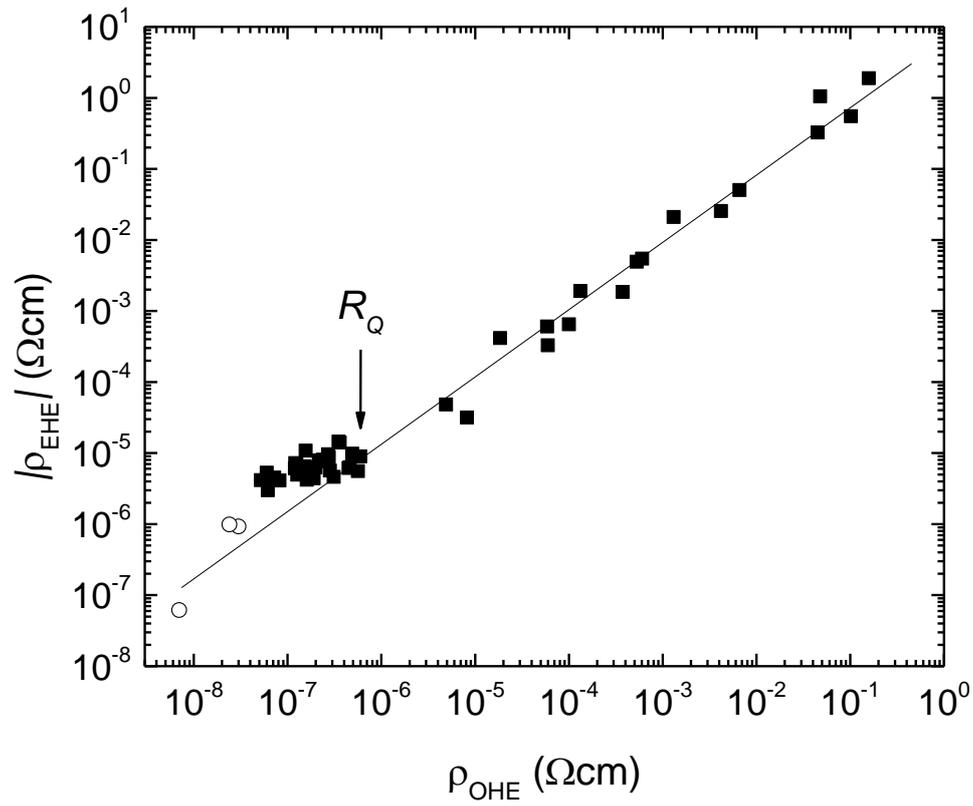

Fig. 5